\documentclass[12pt]{article}

\newenvironment{wileykeywords}{\textsf{Keywords:}\hspace{\stretch{1}}}{\hspace{\stretch{1}}\rule{1ex}{1ex}}

\usepackage{amsmath,amssymb}
\usepackage{graphicx}
\usepackage{color}
\usepackage{dcolumn}
\usepackage{bm}
\usepackage[numbers,super,comma,sort&compress]{natbib}

\definecolor{background-color}{gray}{0.98}

\title{Origin of pressure-induced insulator-to-metal transition in the van der Waals compound FePS$_3$ from first-principles calculations}

\author{Robert A. Evarestov\thanks{Department of Quantum Chemistry, Saint Petersburg State University, 7/9 Universitetskaya Naberezhnaya, St. Petersburg 199034, Russian Federation; E-mail: r.evarestov@spbu.ru}
and 
Alexei Kuzmin\thanks{Institute of Solid State Physics, University of Latvia, Kengaraga street 8, LV-1063, Riga, Latvia; E-mail: a.kuzmin@cfi.lu.lv}}

\begin{document}

\maketitle

\begin{abstract}
Pressure-induced insulator-to-metal transition has been studied in the van der Waals compound iron thiophosphate (FePS$_3$) using first-principles  calculations within the periodic linear combination of atomic orbitals (LCAO) method with hybrid Hartree-Fock-DFT B3LYP functional. 
Our calculations reproduce correctly the insulator-to-metal  transition (IMT) at $\sim$15 GPa, which is accompanied by a reduction of the unit cell volume and of the vdW gap. 
We found from the detailed analysis of the projected density of states that
the 3p states of phosphorus atoms contribute significantly at the bottom of the conduction band. As a result, the collapse of the band gap occurs due to changes in the electronic structure of FePS$_3$ induced by relative displacements of phosphorus or sulfur atoms along the $c$-axis direction under pressure.
\end{abstract}

\begin{wileykeywords}
FePS$_3$, layered compound, high pressure, insulator-to-metal transition, first principles calculations.
\end{wileykeywords}

\clearpage


\begin{figure}[h]
\centering
\colorbox{background-color}{
\fbox{
\begin{minipage}{1.0\textwidth}
\includegraphics[width=50mm,height=50mm]{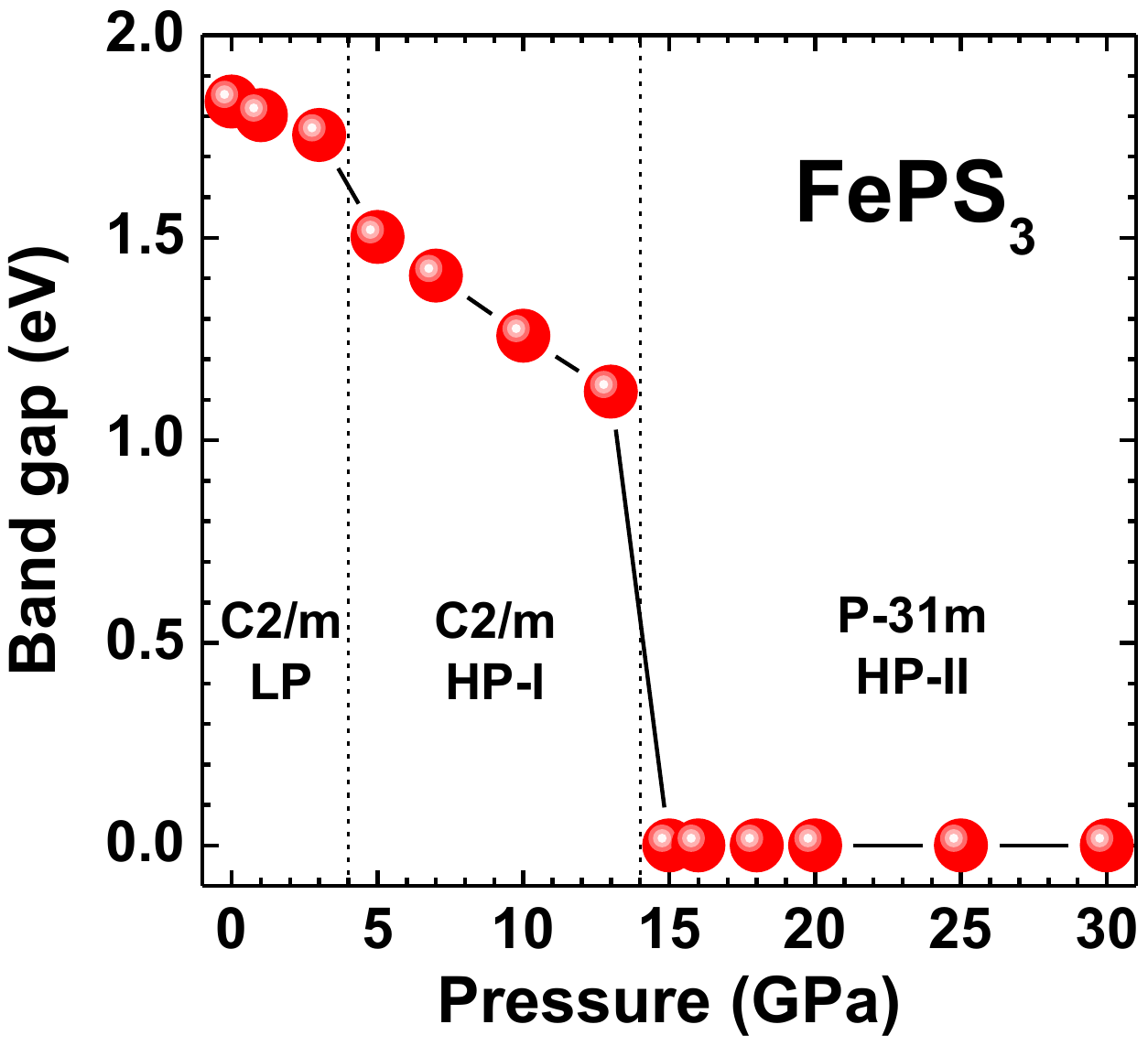} 
\\
Pressure-induced phase transitions in iron thiophosphate (FePS$_3$) at $\sim$4 GPa and $\sim$15 GPa were studied using first-principles calculations. The calculations reproduce the insulator-to-metal transition (IMT) at $\sim$15 GPa, accompanied by a reduction of the unit cell volume and of the van der Waals gap. The origin of the IMT is attributed to the pressure-induced changes in the FePS$_3$ electronic structure caused by the relative displacement of phosphorus or sulfur atoms along the $c$-axis direction.
\end{minipage}
}}
\end{figure}

  \makeatletter
  \renewcommand\@biblabel[1]{#1.}
  \makeatother

\bibliographystyle{apsrev}

\renewcommand{\baselinestretch}{1.5}
\normalsize

\clearpage

\section*{\sffamily \Large INTRODUCTION} 

Iron thiophosphate (FePS$_3$) belongs to a family of van der Waals (vdW) layered materials
and attracted recently much attention due to its remarkable physicochemical and magnetic properties  \cite{Mayorga2017,Wang2018a,Burch2018,Li2019,Gong2019}.
FePS$_3$ is a magnetic semiconductor with the band gap of 0.5-1.6 eV \cite{Brec1979,Foot1980,Haines2018} and 
intrinsic antiferromagnetism below the N\'{e}el temperature of about 120 K \cite{LeFlem1982,Kurosawa1983,Joy1992,Rule2007}. Besides, it can be exfoliated into single and few-layer sheets having enhanced catalytic activity \cite{Cheng2018,Zhu2018}.

At low (ambient) pressure , the crystallographic structure of FePS$_3$ (Fig.\ \ref{fig1}) is composed of  2D layers extended parallel to the $ab$-plane and separated by the vdW gaps along the $c$-axis \cite{Haines2018,Rule2007,Ouvrard1985}.
Each layer contains Fe atoms octahedrally coordinated by six S atoms and P atoms tetrahedrally coordinated by three S atoms and one P atom, forming a [P$_2$S$_6$]$^{4-}$ unit
\cite{Haines2018,Rule2007,Ouvrard1985}. 
The vdW gap (defined as the shortest distance between the adjacent S layers) is 
about  2.8 \AA. 

In low-pressure (LP) phase  \cite{Haines2018}, FePS$_3$ has monoclinic lattice with space group $C2/m$ (No. 12)
and  two FePS$_3$ formula units in the primitive unit cell, but four formula 
units in the crystallographic unit cell. The atoms occupy the following Wyckoff positions: Fe 4g(0,y,0), P 4i(x,0,z), S1 4i(x,0,z), S2 8j(x,y,z). 
The antiferromagnetic ordering in FePS$_3$ is controlled by in-plane interactions between the high-spin ($S$ = 2) Fe$^{2+}$ ions arranged on a honeycomb lattice \cite{LeFlem1982,Joy1992}. 
They are coupled ferromagnetically to the two nearest Fe$^{2+}$ neighbours and antiferromagnetically to the third one. As a result, iron moments form in the $ab$-planes 
ferromagnetic chains coupled antiferromagnetically to each other  
\cite{Kurosawa1983,Joy1992,Lancon2016}. 
The weak vdW interaction between layers results in an Ising-type antiferromagnetic ordering which remains preserved down to the monolayer limit \cite{Lee2016}. 

According to recent pressure-dependent X-ray diffraction experiments \cite{Haines2018,Wang2018}, FePS$_3$ exhibits two phase transitions upon increasing pressure. At about 4 GPa, it transforms to intermediate-pressure phase (HP-I) with monoclinic ($C2/m$ (No. 12)) symmetry. The principal difference between LP and HP-I phases is  a displacement of the unit cell along the $a$-axis in the latter, so that iron ions in the adjacent layers become located on top of each other along the $c$-axis, and the $\beta$ angle between the $a$ and $c$ axes reduces from 107.34$^{\circ}$\ to 89.33$^{\circ}$\  (Fig.\ \ref{fig1} and Table\ \ref{table1}).   The antiferromagnetic ordering survives in the HP-I phase \cite{Wang2018}.

In the high-pressure (HP-II) phase above $\sim$14 GPa, FePS$_3$ crystal belongs to the space group  $P\bar{3}1m$ (No. 162) with the hexagonal lattice and two formula units in the primitive unit cell. The atoms occupy the following Wyckoff positions: Fe 2c(1/3,2/3,0), P 2e(0,0,z), S 6k(x,0,z).  The transition to the HP-II phase is accompanied by 10.6\% volume collapse (due to a significant reduction of the $b$ and $c$ lattice parameters), abrupt spin-crossover transition  from magnetic high-spin ($S$ = 2) to non-magnetic low-spin ($S$ = 0) state, and insulator-to-metal transition (IMT) \cite{Haines2018,Wang2018}.
It was also found that the in-plane metallization makes most contribution to the IMT phenomenon \cite{Wang2018}. 
Note also that the resistivity of FePS$_3$ shows stronger dependence on  pressure than on temperature \cite{Haines2018,Wang2018}. 

To understand the mechanism of the pressure-driven IMT transition in FePS$_3$,  
first-principles calculations based on the plane-wave density functional theory (DFT) 
were recently performed in the range from 0 to 35 GPa in Ref.\ \cite{Zheng2019}. 
The two structural phase transitions were correctly reproduced to occur at about 5 and 17 GPa.
The calculations also showed that the LP and HP-I phases posses antiferromagnetic ordering 
in agreement with the experiment. The band gap of  about 1.31 eV was found in the LP phase. It decreases to 1.00 eV at 10 GPa in the HP-I phase, and, finally, down to zero in the metallic HP-II phase.  
The analysis of the orbital projected density of states allowed the authors to determine the origin of the electronic states above and below the Fermi level. It was concluded that in both LP and HP-I phases, the bottom of the conduction band is formed mainly by  the 3d(Fe) and 3p(S) states, whereas the valence band originates mainly from the 3d(Fe) states \cite{Zheng2019}. In the HP-II phase, the states near the Fermi level are mainly of the 3d(Fe) origin \cite{Zheng2019}.  

In the present study, we concentrate on the detailed understanding of the origin of pressure-induced IMT in FePS$_3$. We demonstrate that opposite to previous work \cite{Zheng2019}, the IMT is determined 
by the significant contribution of 3p(P) states into conduction band bottom, which can be tuned by the relative displacement of phosphorus or sulfur atoms along the $c$-axis. We show that metallic conductivity could occur in any of LP, HP-I, and HP-II phases when P and S atoms are located within one plane upon displacement.

\section*{\sffamily \Large METHODOLOGY}

Pressure-dependent properties of FePS$_3$  have been studied using the first-principle linear combination of atomic orbitals (LCAO) calculations as implemented in the CRYSTAL17 code \cite{crystal17}. 
All-electron triple-zeta valence (TZV) basis sets 
augmented by one set of polarization functions (pob-TZVP) \cite{Peintinger2013} have been employed for Fe, P, and S atoms.

The accuracy in evaluating the Coulomb series and the exchange series was controlled by a set of tolerances, which were taken to be (10$^{-8}$, 10$^{-8}$, 10$^{-8}$, 10$^{-8}$, 10$^{-16}$). The Monkhorst-Pack scheme
\cite{Monkhorst1976} for an 8$\times$8$\times$8 $\textbf{k}$-point mesh in the
Brillouin zone was applied.  The SCF calculations were performed for several hybrid DFT-HF functionals with a 10$^{-10}$ tolerance on change in the total energy.  The best agreement with the experimental structural data  \cite{Haines2018} and 
the band gap value \cite{Brec1979,Foot1980} for the low-pressure ($P$=0 GPa) phase was obtained for Becke's 3-parameter functional (B3LYP-13\%) \cite{B3LYP}, which was used in all reported simulations. The
percentage (13\%) defines the Hartree-Fock admixture in the exchange part of DFT functional. All calculations were performed using a restricted closed-shell hamiltonian, i.e. for non-magnetic structures. 
We believe that such approximation is consistent with the experimental temperature dependence of the electrical resistance  (see Figs.~4 and 5 in Ref.~\cite{Haines2018} and Fig.~3 in Ref.~\cite{Wang2018}), which demonstrates the IMT in FePS$_3$ in a wide range of temperatures up to 300 K, i.e. far above its N\'eel temperature of $T_N$ = 120 K \cite{LeFlem1982,Kurosawa1983,Joy1992,Rule2007}.

The lattice parameters and atomic fractional coordinates were optimized for each selected pressure in the range of 0--30 GPa for three phases (Figs.\ \ref{fig2} and \ref{fig3}):
low-pressure (LP) monoclinic (space group $C2/m$) phase, high-pressure (HP-I) monoclinic (space group $C2/m$) phase, and high-pressure (HP-II) trigonal (space group $P\bar{3}1m$) phase. The starting structural parameters were taken from the experimental data \cite{Haines2018}.   
The structure optimization at required pressure was performed using the approach developed in Ref.\ \cite{Jackson2015}.

The phonon frequencies were computed at the center of the Brillouin zone (the $\Gamma$-point)  within the harmonic approximation using the direct (frozen-phonon) method\cite{crystal17,Pascale2004} for each FePS$_3$ phase. The primitive cell of FePS$_3$ in all phases includes 10 atoms (2 Fe, 2 P, and 6 S), so that 30 phonon modes are expected and are classified as the Raman-active (R), infrared-active (IR), and silent (S) modes.

According to group theoretical analysis for space group $C2/m$ in LP and HP-I phases, there are the 7B$_g$ and 8A$_g$ Raman-active even modes, whereas the 9B$_u$  and 6A$_u$ odd modes are infrared-active (three of them (2B$_u$ and 1A$_u$) are acoustic modes with zero frequency at the  $\Gamma$-point).

In HP-II phase with space group $P\bar{3}1m$, the 5E$_g$ and 3A$_{1g}$ even modes  are 
Raman-active, whereas the 5E$_u$ and 4A$_{2u}$ odd modes are infrared-active (two of them (1E$_u$ and 1A$_{2u}$) are acoustic modes with zero frequency at the  $\Gamma$-point). There are  also three silent modes (1A$_{1u}$ and 2A$_{2g}$) in HP-II phase.

The obtained structural parameters such as lattice parameters ($a$, $b$, $c$, $\beta$) and atomic fractional coordinates  ($x$, $y$, $z$) as well as the values of the band gap $E_g$ are reported in Table\ \ref{table1}.
The phonon frequencies calculated at $P$=0, 10, and 18 GPa  are given in Table\ \ref{table2}.
The pressure dependence of the van der Waals (vdW) gap defined as the distance between two planes containing lowest and highest sulfur atoms in the two neighbouring layers (Fig.\ \ref{fig1}) is reported in Fig.\ \ref{fig3}.
Calculated  band structures and total/projected density of states for the LP, HP-I, and HP-II FePS$_3$ phases are shown in Figs.\ \ref{fig4}, \ref{fig5} and  \ref{fig6}.

Finally, we have performed the calculations of the electronic structure of FePS$_3$ for artificial 
situations with P atoms displaced along the $c$-axis. The crystal structures were fixed at the ones optimized for LP (0 GPa), HP-I (10 GPa), and HP-II (18 GPa) phases, while the displacement of P atoms  $\Delta z$  was varying between $-$0.1 \AA\ and 0.5 \AA. The obtained variations of the band gap  $E_g$ are shown  in Fig.\ \ref{fig7}.

\section*{\sffamily \Large RESULTS AND DISCUSSION}

The structural properties of FePS$_3$ from our LCAO calculations, which correspond to the lowest temperature limit ($T$=0~K), agree with the experimental findings from Ref.\ \cite{Haines2018} (Table\ \ref{table1}). 
The comparison for the phonon frequencies is possible only for the LP phase, 
for which the experimental infrared and Raman spectra measured at room temperature are available \cite{Bernasconi1988,Wang2016}. The calculated values of phonon modes at the  $\Gamma$-point (Table\ \ref{table2}) are in qualitative agreement with the experimental data, however the low-frequency modes are slightly overestimated. 
Our calculations reproduce correctly the transition to metallic state at $\sim$15 GPa (Fig.\ \ref{fig3}), which is accompanied by a reduction of the unit cell volume by $\sim$7\% (Fig.\ \ref{fig2}) and of the vdW gap by $\sim$13\% (Fig.\ \ref{fig3}). 

Pressure dependence of the band gap $E_g$ in LP, HP-I, and HP-II phases was
evaluated from the band structure calculations performed for optimized crystal lattice geometry (lattice parameters and atomic fractional coordinates) and is reported in Fig.\ \ref{fig3}. The results suggest that low-pressure monoclinic $C2/m$ lattice is very stable against the compression, showing no transition to metallic state up to 30 GPa. The monoclinic $C2/m$ lattice of the HP-I phase is 
more pliable, however the collapse of the band gap was only observed for pressures starting from 30 GPa and above. 
Much softer behaviour was found for trigonal $P\bar{3}1m$ lattice, in which the band gap drops from $\sim$1.5 eV to zero in the pressure range from 0 to 15 GPa.

To understand the origin of the transition to metallic state, detailed analysis of the electronic band structure of FePS$_3$ was performed (Fig.\ \ref{fig4}). 
The calculated band gap for the FePS$_3$ LP phase at $P$=0 GPa is 1.84 eV, i.e. the material is an  insulator. Upon increasing pressure to 3 GPa, the band gap reduces monotonically to 1.75 eV. 
The transition to the HP-I phase occurs at $\sim$4 GPa and leads to the abrupt change of the band gap 
down to 1.50 eV. The monotonic decrease of the band gap continues in the HP-I phase, finally achieving 
the value of 1.12 eV at 13 GPa. Note that the band gap is indirect in both LP and HP-I phases. 
Further increase of pressure leads to a collapse of the band gap and a transition to the metallic HP-II phase.   

Our results suggest that the top of the valence bands in the LP and HP-I phases is mainly composed of the 3d(Fe) and 3p(S) states (Figs.\ \ref{fig5} and \ref{fig6}), whereas the bottom of the conduction bands originates mainly from  3d$_{xz,yz,xy}$(Fe), 3p$_{z}$(P) and 3p$_{x}$(S) states. 
Note that in the recent work \cite{Zheng2019} not enough attention has been paid to the contribution of the 3p(P) states to the conduction band. 

A pressure-dependent variation of the calculated electronic structure of FePS$_3$ is due to several effects that complicate its analysis.  
These include the reduction of the vdW gap, a compression of the 2D layers in the $ab$-plane and changes in atomic fractional coordinates. To simplify the task, we considered three artificial models based on the optimized crystallographic structures for the LP (0 GPa), HP-I (10 GPa), and HP-II (18 GPa) phases. 
In these models, the position of P atoms was varying along the $c$-axis direction relative to the optimized one (Figs.\ \ref{fig7} and \ref{fig8}). We found that the displacement of P atoms in the direction of the plane formed by sulfur atoms in the LP and HP-I phases (Fig.\ \ref{fig1}) leads to a decrease of the band gap up to the transition to the metallic state. 

In the HP-II phase at 18 GPa, P atoms have 3-fold triangular coordination by sulfur atoms, thus they are already located within the S atom plane (Fig.\ \ref{fig1}).
Therefore, the displacement of P atoms along the $c$-axis direction moves them away from the plane formed by sulfur atoms that results in the opposite effect (Fig.\ \ref{fig7}), i.e. opening of the band gap when the displacement is large enough ($\Delta z$(P)$>$0.1 \AA). 

Thus, hybridization of the 3d(Fe), 3p(P), and 3p(S) states around the Fermi level plays an important role in the electronic structure of FePS$_3$. 
Under increasing pressure, the relative displacement of the P atoms leads to the broadening of both valence and conduction bands that results in the band gap collapse, i.e. insulator-to-metal transition. The control over 
the relative displacements of P atoms can be used to tune the transition.

\section*{\sffamily \Large CONCLUSIONS}

First-principles LCAO calculations using hybrid DFT-HF B3LYP functional have been performed to understand the insulator-to-metal transition in FePS$_3$. The calculated insulator-to-metal  transition occurs at $\sim$15 GPa and is accompanied by the unit cell volume and van der Waals gap reduction and the space group change from monoclinic  $C2/m$ to trigonal $P\bar{3}1m$. The obtained results are in agreement with the available experimental data \cite{Haines2018} and  recent calculations \cite{Zheng2019}.

The origin of the insulator-to-metal transition is attributed by us to the pressure-induced broadening of valence and conduction bands in the FePS$_3$ electronic structure caused by the relative displacement of phosphorus or sulfur atoms along the $c$-axis direction. This displacement leads to the P and S atoms arrangement within one plane. Our calculations show (Fig.\ \ref{fig7}) that even in the absence of the vdW gap and lattice parameter reduction due to compression,  a sufficiently large displacement of P atoms  could lead to metallic conductivity in both LH and HP-I phases, whereas could produce opposite effect, i.e. opening of the band gap, in the HP-II phase. Such behaviour is explained by the significant contribution of the 3p(P) states in the conduction band bottom (Figs.\ \ref{fig5} and \ref{fig7}).

\subsection*{\sffamily \large ACKNOWLEDGMENTS}
The authors acknowledge the assistance of the University Computer
Center of Saint-Petersburg State University in the accomplishment
of high-performance computations.
A.K. is grateful to the Latvian Council of Science project no. lzp-2018/2-0353 for financial support.


\clearpage



\clearpage

\begin{figure}
	\centering
	\caption{ Crystallographic structure  of FePS$_3$ in the low-pressure ($P$=0 \& 30 GPa) monoclinic (space group $C2/m$) phase, intermediate pressure ($P$=10 \& 30 GPa) monoclinic (space group $C2/m$) phase and high-pressure ($P$=18 \& 30 GPa) trigonal (space group $P\bar{3}1m$) phase. The van der Waals (vdW) gaps are indicated. The illustrations were created using the VESTA software \protect\cite{VESTA}.}
	\label{fig1}
\end{figure}

\begin{figure}
	\centering
	\caption{Pressure dependence of the calculated lattice parameters and primitive cell volume in FePS$_3$.   }
	\label{fig2}
\end{figure}

\begin{figure}
	\centering
	\caption{Pressure dependence of the calculated band gap $E_g$ and the van der Waals (vdW) gap in FePS$_3$.   }
	\label{fig3}
\end{figure}

\begin{figure*}
	\centering
	\caption{Band structure diagram for the LP, HP-I, and HP-II FePS$_3$ phases.
		The energy zero is set at the top of the valence band (Fermi energy position). }
	\label{fig4}
\end{figure*}

\begin{figure*}
	\centering
	\caption{Total and projected density of states (DOS) for the LP, HP-I, and HP-II FePS$_3$ phases. The energy zero is set at the top of the valence band (Fermi energy position). }
	\label{fig5}
\end{figure*}

\begin{figure*}
	\centering
	\caption{Total and projected onto the set of atomic orbitals density of states (DOS) for the LP, HP-I, and HP-II FePS$_3$ phases. The energy zero is set at the top of the valence band (Fermi energy position). }
	\label{fig6}
\end{figure*}

\begin{figure}
	\centering
	\caption{Dependence of the band gap $E_g$ in LP, HP-I, and HP-II FePS$_3$ phases on the displacement of phosphorus atoms $\Delta z$(P) along the $c$-axis.  }
	\label{fig7}
\end{figure}

\begin{figure*}
	\centering
	\caption{Total and projected density of states (DOS) for the LP ($C2/m$) FePS$_3$ phase as a function of phosphorus atoms displacement $\Delta z$(P) along the $c$-axis. The energy zero is set at the top of the valence band (Fermi energy position). }
	\label{fig8}
\end{figure*}


\clearpage

\begin{center}
	\includegraphics[width=0.9\columnwidth,keepaspectratio=true]{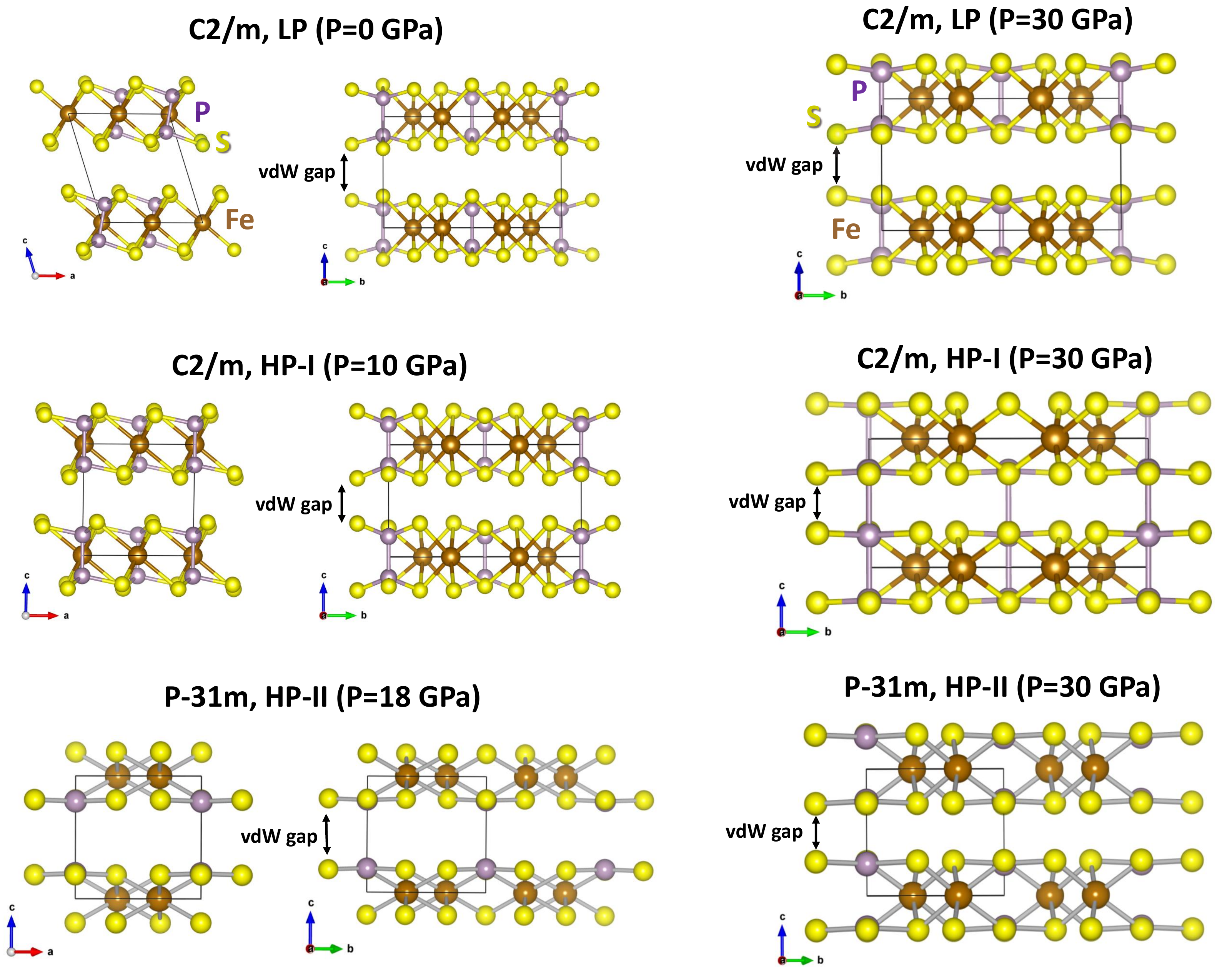}
\end{center}
\vspace{0.25in}
\hspace*{3in}
{\Large
	\begin{minipage}[t]{3in}
		\baselineskip = .5\baselineskip
		Figure 1 \\
		R. A. Evarestov, A. Kuzmin \\
		J.\ Comput.\ Chem.
	\end{minipage}
}

\clearpage

\begin{center}
	\includegraphics[width=0.6\columnwidth,keepaspectratio=true]{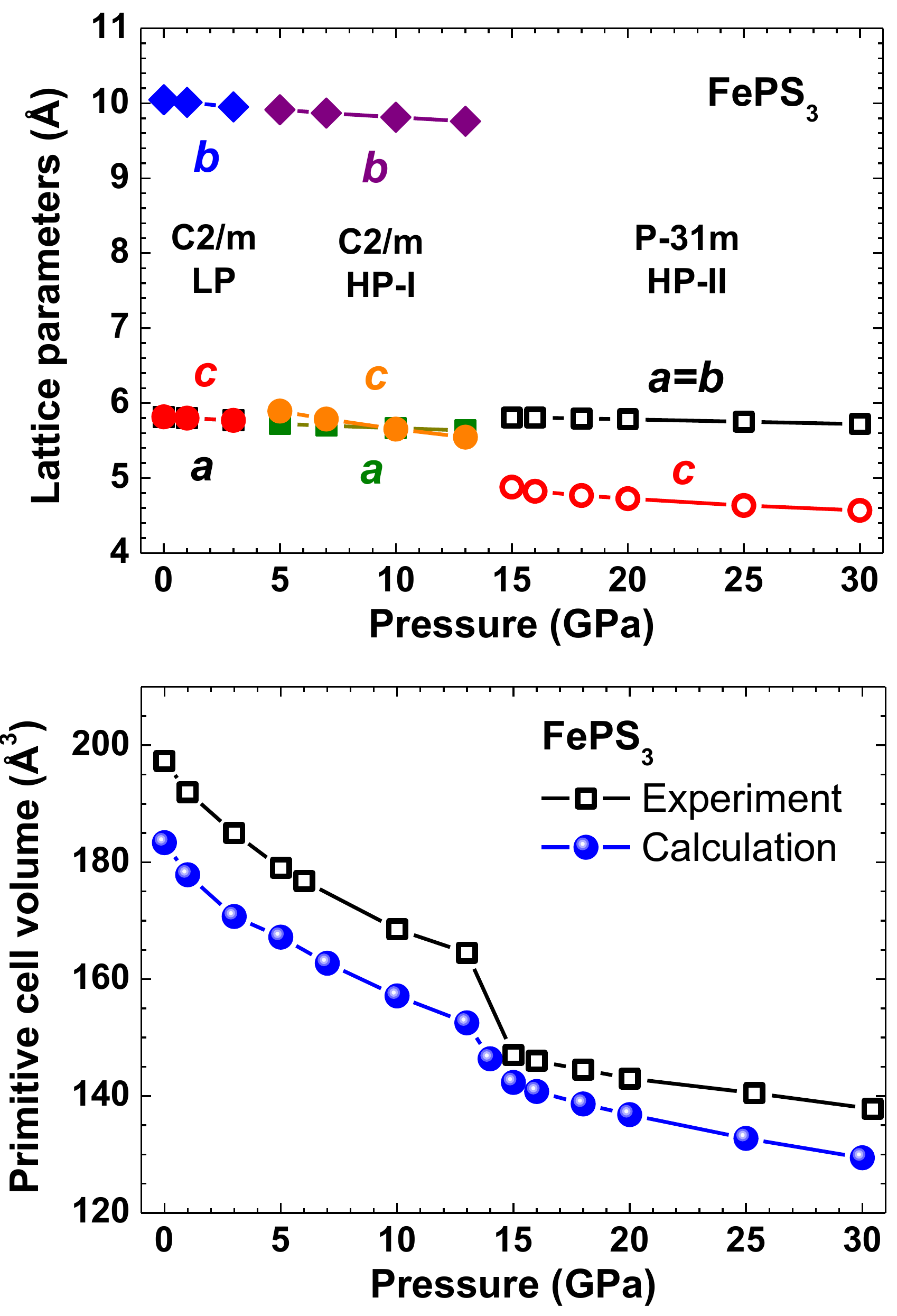}
\end{center}
\vspace{0.25in}
\hspace*{3in}
{\Large
	\begin{minipage}[t]{3in}
		\baselineskip = .5\baselineskip
		Figure 2 \\
		R. A. Evarestov, A. Kuzmin \\
		J.\ Comput.\ Chem.
	\end{minipage}
}

\clearpage

\begin{center}
	\includegraphics[width=0.6\columnwidth,keepaspectratio=true]{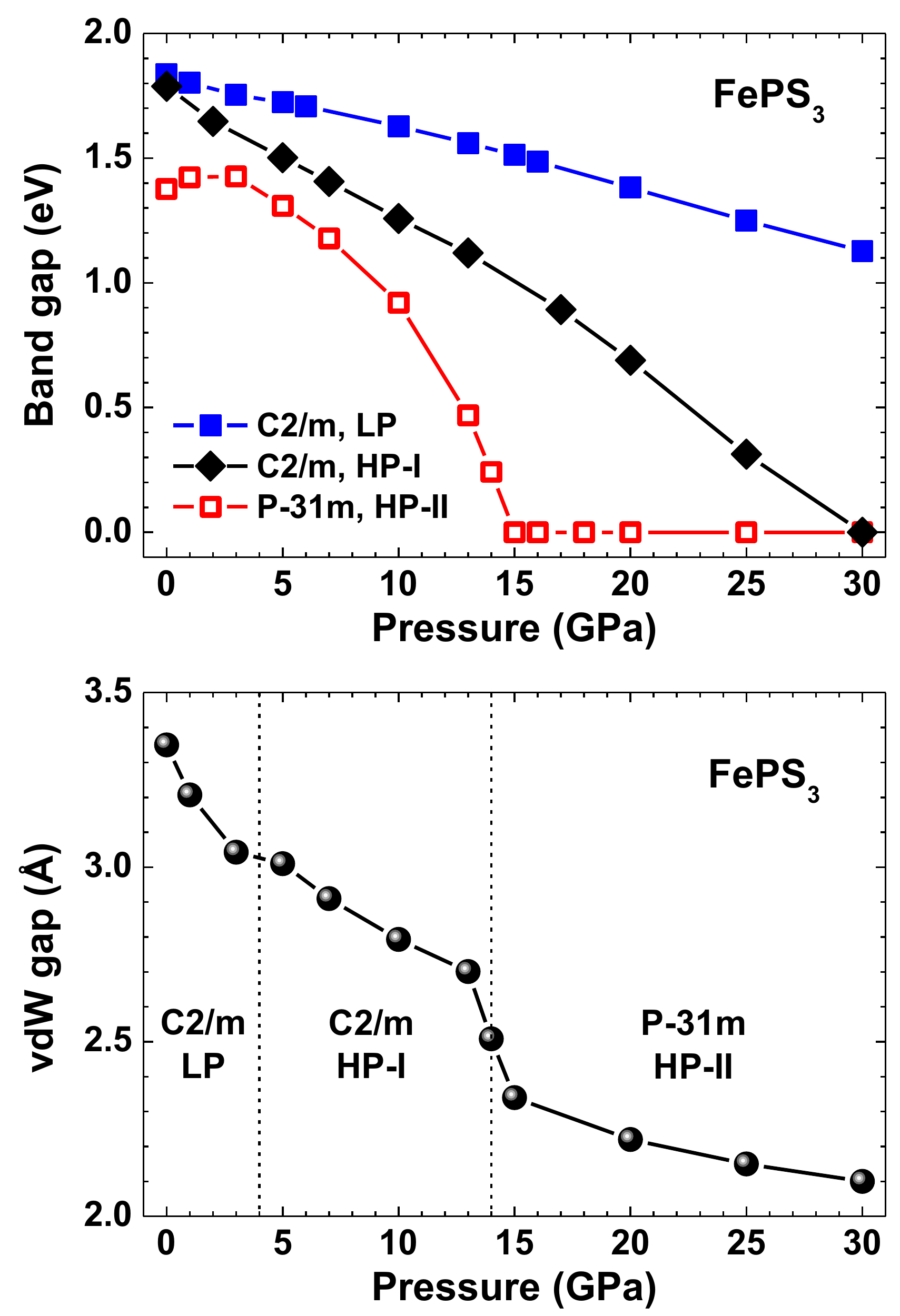}
\end{center}
\vspace{0.25in}
\hspace*{3in}
{\Large
	\begin{minipage}[t]{3in}
		\baselineskip = .5\baselineskip
		Figure 3 \\
		R. A. Evarestov, A. Kuzmin \\
		J.\ Comput.\ Chem.
	\end{minipage}
}

\clearpage

\begin{center}
	\includegraphics[width=0.95\columnwidth,keepaspectratio=true]{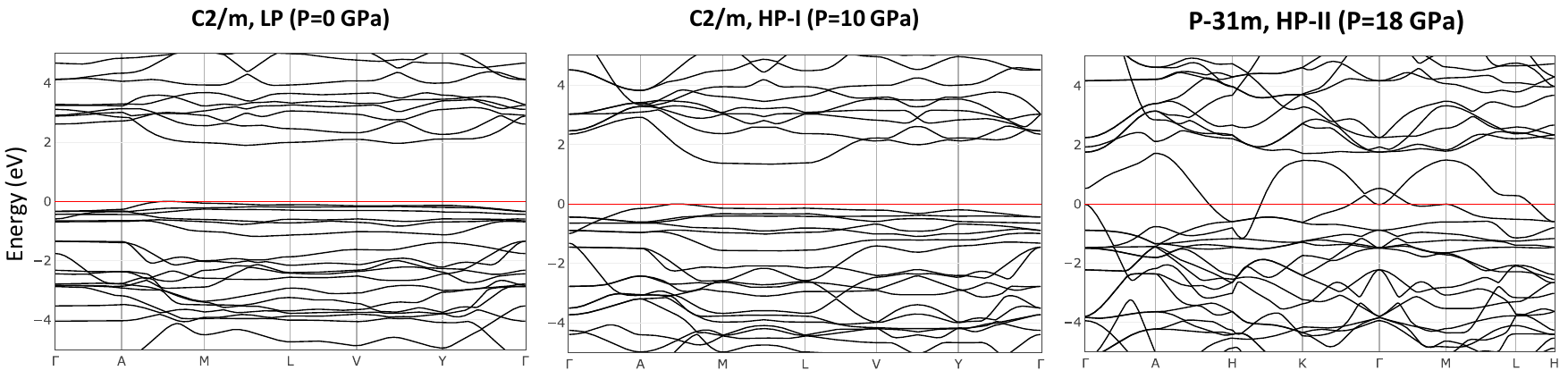}
\end{center}
\vspace{0.25in}
\hspace*{3in}
{\Large
	\begin{minipage}[t]{3in}
		\baselineskip = .5\baselineskip
		Figure 4 \\
		R. A. Evarestov, A. Kuzmin \\
		J.\ Comput.\ Chem.
	\end{minipage}
}

\clearpage

\begin{center}
	\includegraphics[width=0.95\columnwidth,keepaspectratio=true]{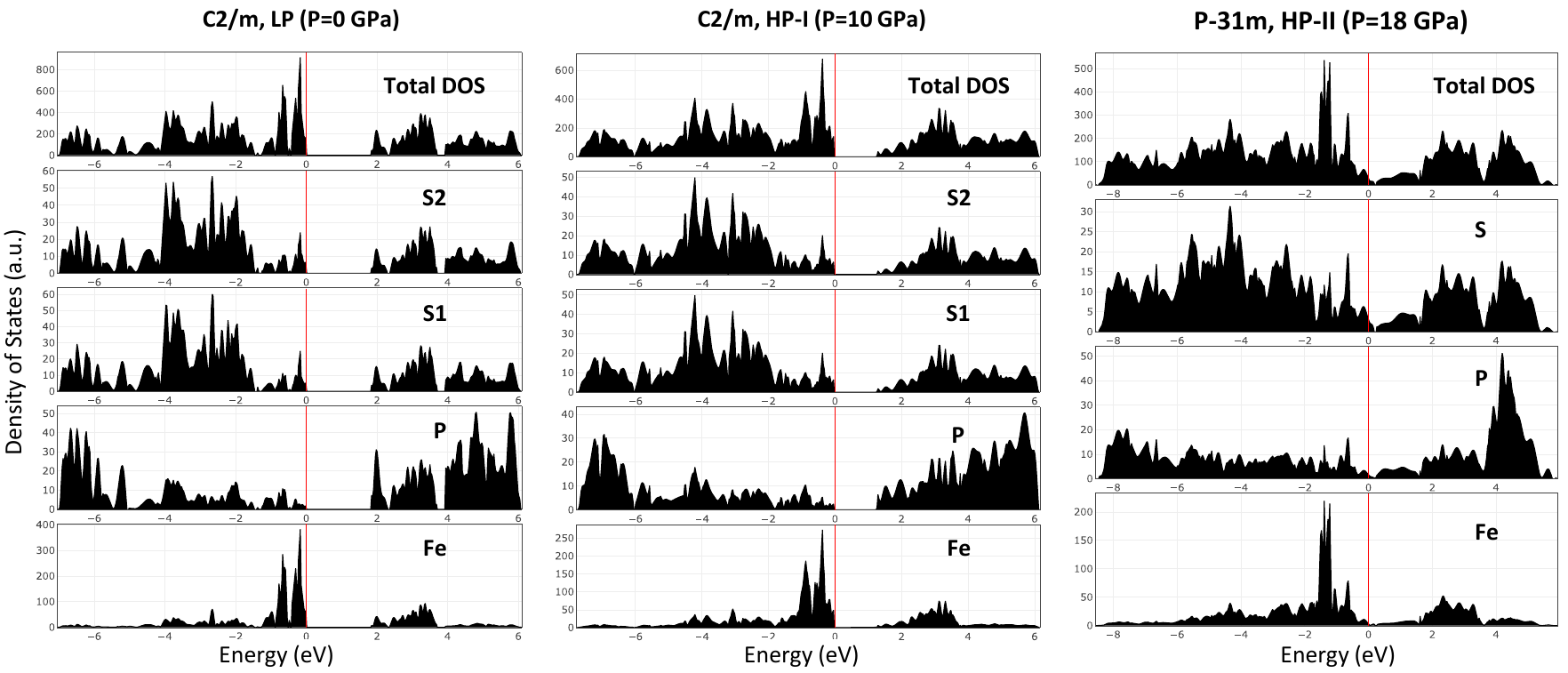}
\end{center}
\vspace{0.25in}
\hspace*{3in}
{\Large
	\begin{minipage}[t]{3in}
		\baselineskip = .5\baselineskip
		Figure 5 \\
		R. A. Evarestov, A. Kuzmin \\
		J.\ Comput.\ Chem.
	\end{minipage}
}

\clearpage

\begin{center}
	\includegraphics[width=0.85\columnwidth,keepaspectratio=true]{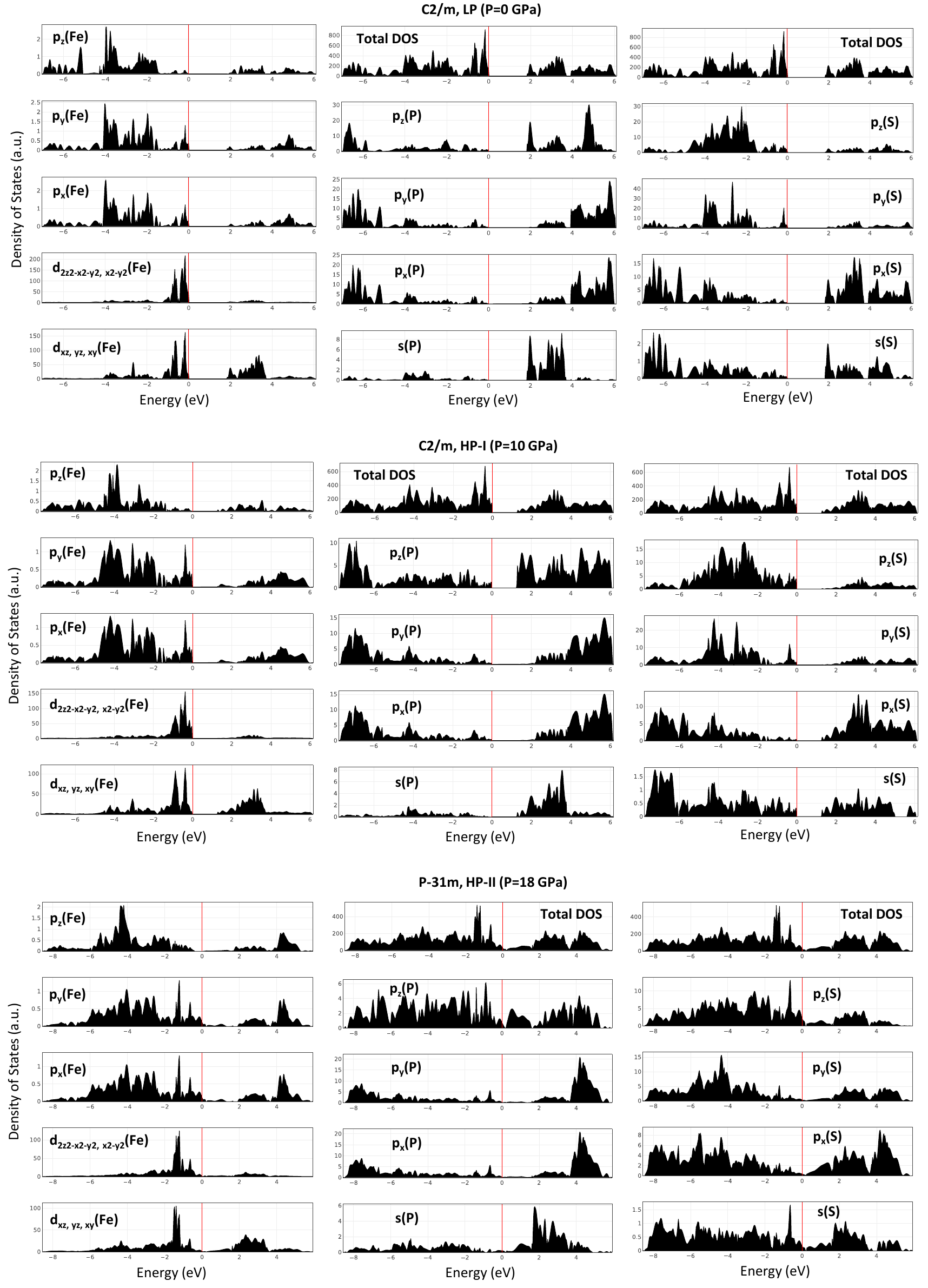}
\end{center}
\vspace{0.25in}
\hspace*{3in}
{\Large
	\begin{minipage}[t]{3in}
		\baselineskip = .5\baselineskip
		Figure 6 \\
		R. A. Evarestov, A. Kuzmin \\
		J.\ Comput.\ Chem.
	\end{minipage}
}

\clearpage

\begin{center}
	\includegraphics[width=0.6\columnwidth,keepaspectratio=true]{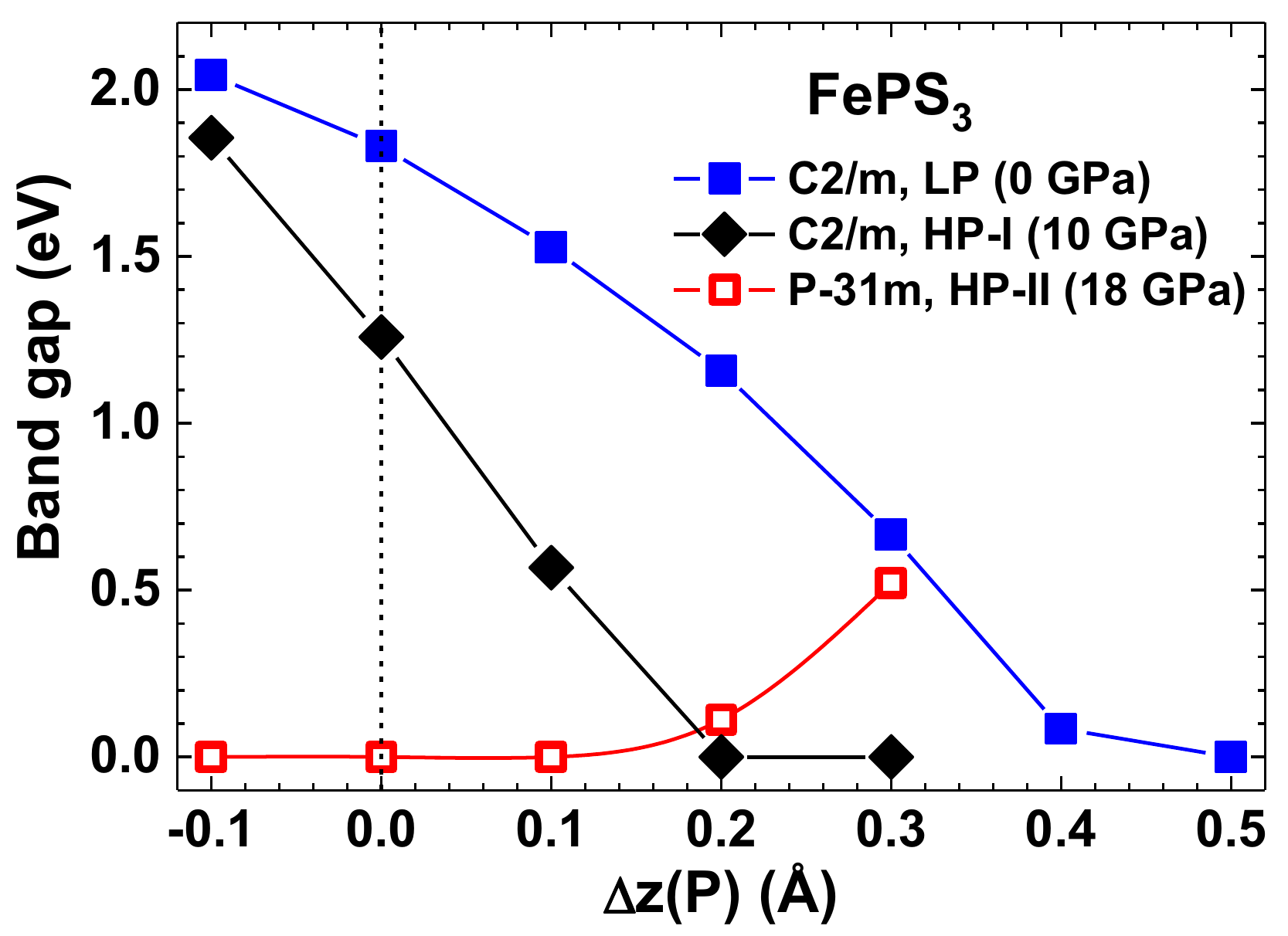}
\end{center}
\vspace{0.25in}
\hspace*{3in}
{\Large
	\begin{minipage}[t]{3in}
		\baselineskip = .5\baselineskip
		Figure 7 \\
		R. A. Evarestov, A. Kuzmin \\
		J.\ Comput.\ Chem.
	\end{minipage}
}

\clearpage

\begin{center}
	\includegraphics[width=0.95\columnwidth,keepaspectratio=true]{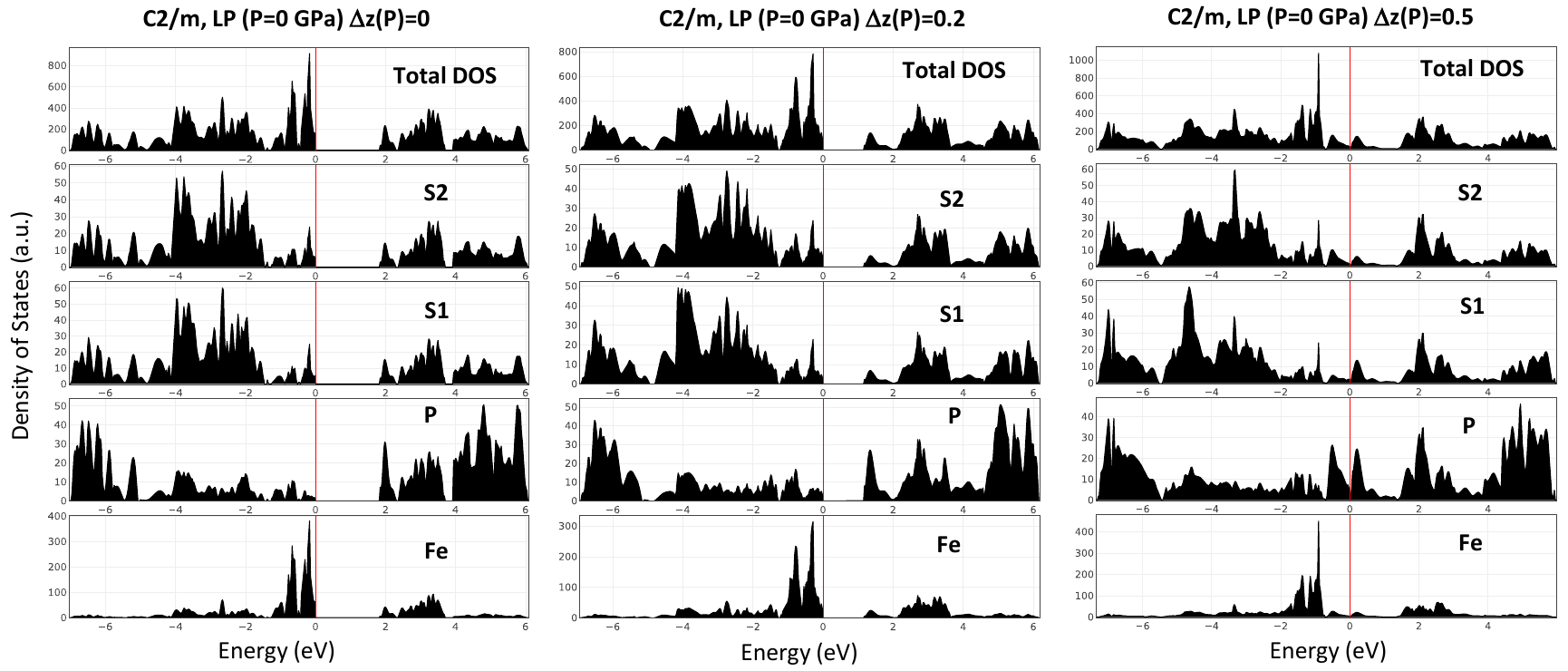}
\end{center}
\vspace{0.25in}
\hspace*{3in}
{\Large
	\begin{minipage}[t]{3in}
		\baselineskip = .5\baselineskip
		Figure 8 \\
		R. A. Evarestov, A. Kuzmin \\
		J.\ Comput.\ Chem.
	\end{minipage}
}

\clearpage


\begin{table*}
	\footnotesize
	\caption{Crystallographic parameters and band gap values  for FePS$_3$ at 0, 10, and 18 GPa. Experimental data are taken from Refs. \protect\cite{Brec1979,Foot1980,Haines2018}.  }
	\label{table1}     
	\centering 
	\renewcommand{\arraystretch}{0.8}
	\begin{tabular}{lllllll} 
		\\
		\hline 
		& \multicolumn{2}{c}{Space group $C2/m$ (12)}  &  \multicolumn{2}{c}{Space group $C2/m$ (12)} & \multicolumn{2}{c}{Space group $P\bar{3}1m$ (162)}  \\  
		& \multicolumn{2}{c}{LP ($P$=0 GPa)}   & \multicolumn{2}{c}{HP-I ($P$=10 GPa)}   &\multicolumn{2}{c}{HP-II ($P$=18 GPa)}  \\ 
		
		& Experiment \protect\cite{Haines2018} & LCAO    & Experiment \protect\cite{Haines2018} & LCAO       & Experiment \protect\cite{Haines2018} & LCAO  \\        
		\hline
		a (\AA)  & 5.9428  &  5.816 &  5.7620  & 5.666   & 5.699 &5.791   \\
		b (\AA)  & 10.299 & 10.047  &  9.988  &  9.813  &    &       \\
		c (\AA)  & 6.7160 & 6.600  &  5.803  & 5.652   & 4.818 & 4.786   \\
		
		$\beta$ ($^\circ$) & 107.34 & 108.05 & 89.33   & 90.01  & & \\
		
		y(Fe)  & 0.3320 & 0.3332 & 0.3225   & 0.3333  &    &  \\
		
		x(P)   & 0.0860 & 0.0619 & 0.0   & 0.0  &  &  \\
		z(P)   & 0.1670 & 0.1716 & 0.184   & 0.1868   &-0.206  &-0.2603   \\
		
		x(S1)  & 0.7600 & 0.7251 & 0.638   & 0.6375 & 0.3241  & 0.3686   \\
		z(S1)  & 0.2860 & 0.2332 & 0.259   & 0.2529 & -0.195 & -0.2632  \\
		
		x(S2) &0.2690 &  0.2610 & 0.127   & 0.1813 &  &  \\
		y(S2) &0.1745 &  0.1797 & 0.1624  & 0.1813 &  &  \\
		z(S2) &0.2470 &  0.2308 & 0.299   & 0.2529 &  &  \\ 
		$E_g$ (eV) & 1.5 \protect\cite{Brec1979}      &  1.8 &  & 1.3 & 0 & 0 \\
		& 1.6 \protect\cite{Foot1980}      &      &  &     &   &   \\
		& 0.452 \protect\cite{Haines2018}  &      &  &     &   &   \\
		\hline
	\end{tabular}
\end{table*}   

\clearpage

\begin{table*}
	\footnotesize
	\caption{Calculated phonon frequencies (in cm$^{-1}$) at the  $\Gamma$-point for FePS$_3$ at $P$=0, 10, and 18 GPa. Raman active (R), infrared active (IR), and silent (S) modes are indicated. The acoustic modes with zero frequency are
	not given. The experimental infrared and Raman frequencies (exp.) measured at room temperature \protect\cite{Bernasconi1988,Wang2016} are also reported for comparison. }
	\label{table2}     
	\centering 
	\renewcommand{\arraystretch}{0.8}
	\begin{tabular}{llllllllll} 
		\\
		\hline
		\multicolumn{4}{c}{Space group $C2/m$ (12)}  & \multicolumn{3}{c}{Space group $C2/m$ (12)}  & \multicolumn{3}{c}{Space group $P\bar{3}1m$ (162)}  \\  
		\multicolumn{4}{c}{LP ($P$=0 GPa)}   &   \multicolumn{3}{c}{HP-I ($P$=10 GPa)}  & \multicolumn{3}{c}{HP-II ($P$=18 GPa)}  \\ 
		Mode & Frequency &  Frequency (exp.)& Activity  & Mode & Frequency & Activity  &  Mode & Frequency & Activity   \\    
		\hline
		B$_g$ & 151 & 101 &   R            &B$_g$   &   133     &    R        &     A$_{2u}$      &     168       &       IR    \\
		B$_u$ & 174 &151 &   IR           &B$_u$    &  160     &   IR         &     A$_{2g}$      &     217       &       S     \\
		A$_g$ & 185 &   &   R            &A$_g$    &  204     &  R          &     E$_g$         &      259      &       R     \\
		B$_g$ & 185 &   &   R            &B$_g$    &  204     &  R          &                  &                      &          \\
		B$_u$ & 207 &   &   IR           &A$_u$   &    233     &   IR         &      E$_u$        &    266         &      IR    \\
		A$_u$ & 215 &  &   IR           &B$_u$   &    233     &   IR         &                  &                      &                         \\
		B$_g$ & 239 &  153    &   R            &B$_g$   &   241     &  R          &      E$_g$   &     276      &      R       \\
		A$_g$ & 240 &   &   R            &A$_g$   &  241     &  R          &                  &                      &                         \\
		B$_u$ & 241 &185  &   IR           &B$_u$   &  267     &   IR         &     E$_u$      &     301       &     IR       \\ 
		B$_u$ & 242 &  &     IR         &A$_u$  &    267     &   IR         &                  &                      &                          \\
		A$_u$ & 245 &  &     IR         &B$_u$  &    278     &   IR         &    A$_{2u}$      &      334      &     IR         \\ 
		A$_g$ & 267 & 220  &    R           &A$_g$   &  312     &  R          &  A$_{1g}$      &   334      &   R       \\
		A$_u$ & 270 &  &    IR          &A$_g$  &    312     &  R          &      E$_g$        &      352      &      R         \\
		B$_g$ & 272 &   &    R           &B$_g$   &   313     &  R          &                  &                     &                            \\
		A$_g$ & 276 & 244  &    R           &A$_u$  &    313     &   IR         &  E$_g$       &      380      &      R          \\
		B$_g$ & 281 & 277  &   R            &B$_g$   &   323     &  R          &  A$_{2u}$     &      380       &      IR        \\
		B$_g$ & 314 &    &   R            &A$_g$   &  344     &  R          &                  &                        &                             \\
		A$_g$ & 315 &   &    R           &B$_g$   &  344     &  R          &      A$_{1g}$     &       381      &     R           \\
		A$_u$ & 318 &258 &     IR         &A$_u$  &   360     &   IR         &    A$_{1u}$     &       386     &    S            \\
		B$_u$ & 321 &295 &    IR          &B$_u$   &  360     &   IR         &     A$_{2g}$    &     397       &     S      \\
		A$_g$ & 363 & 378 &     R          &A$_g$   &  398     &  R          &     E$_u$       &     401  &     IR  \\
		B$_u$ & 413 &445 &      IR        &B$_u$   &  432     &   IR         &          &          &               \\
		A$_g$ & 506 &  &    R           &B$_g$   &  539     &  R          &         A$_{1g}$  &    416       &      R           \\
		B$_g$ & 508 &  &    R           &A$_g$   &  539     &  R          &     E$_g$      &     486       &  R      \\
		B$_u$ & 526 &  &      IR        &A$_u$  &    565     &   IR         &                    &                   &           \\
		A$_u$ & 530 &578 &      IR        &B$_u$  &    565     &   IR         &    E$_u$     &    532       & IR      \\
		A$_g$ & 574 & 573 &    R           &A$_g$   &  613     &  R          &                &                   &        \\  
		\hline
	\end{tabular}
\end{table*}


\begin{thebibliography}{28}
	\expandafter\ifx\csname natexlab\endcsname\relax\def\natexlab#1{#1}\fi
	\expandafter\ifx\csname bibnamefont\endcsname\relax
	\def\bibnamefont#1{#1}\fi
	\expandafter\ifx\csname bibfnamefont\endcsname\relax
	\def\bibfnamefont#1{#1}\fi
	\expandafter\ifx\csname citenamefont\endcsname\relax
	\def\citenamefont#1{#1}\fi
	\expandafter\ifx\csname url\endcsname\relax
	\def\url#1{\texttt{#1}}\fi
	\expandafter\ifx\csname urlprefix\endcsname\relax\def\urlprefix{URL }\fi
	\providecommand{\bibinfo}[2]{#2}
	\providecommand{\eprint}[2][]{\url{#2}}
	
	\bibitem[{\citenamefont{Mayorga-Martinez
			et~al.}(2017)\citenamefont{Mayorga-Martinez, Sofer, Sedmidubsk\'{y}, Huber,
			Eng, and Pumera}}]{Mayorga2017}
	\bibinfo{author}{\bibfnamefont{C.~C.} \bibnamefont{Mayorga-Martinez}},
	\bibinfo{author}{\bibfnamefont{Z.}~\bibnamefont{Sofer}},
	\bibinfo{author}{\bibfnamefont{D.}~\bibnamefont{Sedmidubsk\'{y}}},
	\bibinfo{author}{\bibfnamefont{{\v{S}}.}~\bibnamefont{Huber}},
	\bibinfo{author}{\bibfnamefont{A.~Y.~S.} \bibnamefont{Eng}},
	\bibnamefont{and} \bibinfo{author}{\bibfnamefont{M.}~\bibnamefont{Pumera}},
	\bibinfo{journal}{ACS Appl. Mater. Interfaces} \textbf{\bibinfo{volume}{9}},
	\bibinfo{pages}{12563} (\bibinfo{year}{2017}).
	
	\bibitem[{\citenamefont{Wang et~al.}(2018{\natexlab{a}})\citenamefont{Wang,
			Shifa, Yu, He, Liu, Wang, Wang, Zhan, Lou, Xia et~al.}}]{Wang2018a}
	\bibinfo{author}{\bibfnamefont{F.}~\bibnamefont{Wang}},
	\bibinfo{author}{\bibfnamefont{T.~A.} \bibnamefont{Shifa}},
	\bibinfo{author}{\bibfnamefont{P.}~\bibnamefont{Yu}},
	\bibinfo{author}{\bibfnamefont{P.}~\bibnamefont{He}},
	\bibinfo{author}{\bibfnamefont{Y.}~\bibnamefont{Liu}},
	\bibinfo{author}{\bibfnamefont{F.}~\bibnamefont{Wang}},
	\bibinfo{author}{\bibfnamefont{Z.}~\bibnamefont{Wang}},
	\bibinfo{author}{\bibfnamefont{X.}~\bibnamefont{Zhan}},
	\bibinfo{author}{\bibfnamefont{X.}~\bibnamefont{Lou}},
	\bibinfo{author}{\bibfnamefont{F.}~\bibnamefont{Xia}}, \bibnamefont{et~al.},
	\bibinfo{journal}{Adv. Func. Mater.} \textbf{\bibinfo{volume}{28}},
	\bibinfo{pages}{1802151} (\bibinfo{year}{2018}{\natexlab{a}}).
	
	\bibitem[{\citenamefont{Burch et~al.}(2018)\citenamefont{Burch, Mandrus, and
			Park}}]{Burch2018}
	\bibinfo{author}{\bibfnamefont{K.}~\bibnamefont{Burch}},
	\bibinfo{author}{\bibfnamefont{D.}~\bibnamefont{Mandrus}}, \bibnamefont{and}
	\bibinfo{author}{\bibfnamefont{J.}~\bibnamefont{Park}},
	\bibinfo{journal}{Nature} \textbf{\bibinfo{volume}{563}}, \bibinfo{pages}{47}
	(\bibinfo{year}{2018}).
	
	\bibitem[{\citenamefont{Li et~al.}(2019)\citenamefont{Li, Ruan, and
			Zeng}}]{Li2019}
	\bibinfo{author}{\bibfnamefont{H.}~\bibnamefont{Li}},
	\bibinfo{author}{\bibfnamefont{S.}~\bibnamefont{Ruan}}, \bibnamefont{and}
	\bibinfo{author}{\bibfnamefont{Y.-J.} \bibnamefont{Zeng}},
	\bibinfo{journal}{Adv. Mater.} \textbf{\bibinfo{volume}{31}},
	\bibinfo{pages}{1900065} (\bibinfo{year}{2019}).
	
	\bibitem[{\citenamefont{Gong and Zhang}(2019)}]{Gong2019}
	\bibinfo{author}{\bibfnamefont{C.}~\bibnamefont{Gong}} \bibnamefont{and}
	\bibinfo{author}{\bibfnamefont{X.}~\bibnamefont{Zhang}},
	\bibinfo{journal}{Science} \textbf{\bibinfo{volume}{363}},
	\bibinfo{pages}{eaav4450} (\bibinfo{year}{2019}).
	
	\bibitem[{\citenamefont{Brec et~al.}(1979)\citenamefont{Brec, Schleich,
			Ouvrard, Louisy, and Rouxel}}]{Brec1979}
	\bibinfo{author}{\bibfnamefont{R.}~\bibnamefont{Brec}},
	\bibinfo{author}{\bibfnamefont{D.~M.} \bibnamefont{Schleich}},
	\bibinfo{author}{\bibfnamefont{G.}~\bibnamefont{Ouvrard}},
	\bibinfo{author}{\bibfnamefont{A.}~\bibnamefont{Louisy}}, \bibnamefont{and}
	\bibinfo{author}{\bibfnamefont{J.}~\bibnamefont{Rouxel}},
	\bibinfo{journal}{Inorg. Chem.} \textbf{\bibinfo{volume}{18}},
	\bibinfo{pages}{1814} (\bibinfo{year}{1979}).
	
	\bibitem[{\citenamefont{Foot et~al.}(1980)\citenamefont{Foot, Suradi, and
			Lee}}]{Foot1980}
	\bibinfo{author}{\bibfnamefont{P.}~\bibnamefont{Foot}},
	\bibinfo{author}{\bibfnamefont{J.}~\bibnamefont{Suradi}}, \bibnamefont{and}
	\bibinfo{author}{\bibfnamefont{P.}~\bibnamefont{Lee}},
	\bibinfo{journal}{Mater. Res. Bulletin} \textbf{\bibinfo{volume}{15}},
	\bibinfo{pages}{189} (\bibinfo{year}{1980}).
	
	\bibitem[{\citenamefont{Haines et~al.}(2018)\citenamefont{Haines, Coak, Wildes,
			Lampronti, Liu, Nahai-Williamson, Hamidov, Daisenberger, and
			Saxena}}]{Haines2018}
	\bibinfo{author}{\bibfnamefont{C.~R.~S.} \bibnamefont{Haines}},
	\bibinfo{author}{\bibfnamefont{M.~J.} \bibnamefont{Coak}},
	\bibinfo{author}{\bibfnamefont{A.~R.} \bibnamefont{Wildes}},
	\bibinfo{author}{\bibfnamefont{G.~I.} \bibnamefont{Lampronti}},
	\bibinfo{author}{\bibfnamefont{C.}~\bibnamefont{Liu}},
	\bibinfo{author}{\bibfnamefont{P.}~\bibnamefont{Nahai-Williamson}},
	\bibinfo{author}{\bibfnamefont{H.}~\bibnamefont{Hamidov}},
	\bibinfo{author}{\bibfnamefont{D.}~\bibnamefont{Daisenberger}},
	\bibnamefont{and} \bibinfo{author}{\bibfnamefont{S.~S.}
		\bibnamefont{Saxena}}, \bibinfo{journal}{Phys. Rev. Lett.}
	\textbf{\bibinfo{volume}{121}}, \bibinfo{pages}{266801}
	(\bibinfo{year}{2018}).
	
	\bibitem[{\citenamefont{{Le Flem} et~al.}(1982)\citenamefont{{Le Flem}, Brec,
			Ouvard, Louisy, and Segransan}}]{LeFlem1982}
	\bibinfo{author}{\bibfnamefont{G.}~\bibnamefont{{Le Flem}}},
	\bibinfo{author}{\bibfnamefont{R.}~\bibnamefont{Brec}},
	\bibinfo{author}{\bibfnamefont{G.}~\bibnamefont{Ouvard}},
	\bibinfo{author}{\bibfnamefont{A.}~\bibnamefont{Louisy}}, \bibnamefont{and}
	\bibinfo{author}{\bibfnamefont{P.}~\bibnamefont{Segransan}},
	\bibinfo{journal}{J. Phys. Chem. Solids} \textbf{\bibinfo{volume}{43}},
	\bibinfo{pages}{455} (\bibinfo{year}{1982}).
	
	\bibitem[{\citenamefont{Kurosawa et~al.}(1983)\citenamefont{Kurosawa, Saito,
			and Yamaguchi}}]{Kurosawa1983}
	\bibinfo{author}{\bibfnamefont{K.}~\bibnamefont{Kurosawa}},
	\bibinfo{author}{\bibfnamefont{S.}~\bibnamefont{Saito}}, \bibnamefont{and}
	\bibinfo{author}{\bibfnamefont{Y.}~\bibnamefont{Yamaguchi}},
	\bibinfo{journal}{J. Phys. Soc. Jap.} \textbf{\bibinfo{volume}{52}},
	\bibinfo{pages}{3919} (\bibinfo{year}{1983}).
	
	\bibitem[{\citenamefont{Joy and Vasudevan}(1992)}]{Joy1992}
	\bibinfo{author}{\bibfnamefont{P.~A.} \bibnamefont{Joy}} \bibnamefont{and}
	\bibinfo{author}{\bibfnamefont{S.}~\bibnamefont{Vasudevan}},
	\bibinfo{journal}{Phys. Rev. B} \textbf{\bibinfo{volume}{46}},
	\bibinfo{pages}{5425} (\bibinfo{year}{1992}).
	
	\bibitem[{\citenamefont{Rule et~al.}(2007)\citenamefont{Rule, McIntyre,
			Kennedy, and Hicks}}]{Rule2007}
	\bibinfo{author}{\bibfnamefont{K.~C.} \bibnamefont{Rule}},
	\bibinfo{author}{\bibfnamefont{G.~J.} \bibnamefont{McIntyre}},
	\bibinfo{author}{\bibfnamefont{S.~J.} \bibnamefont{Kennedy}},
	\bibnamefont{and} \bibinfo{author}{\bibfnamefont{T.~J.} \bibnamefont{Hicks}},
	\bibinfo{journal}{Phys. Rev. B} \textbf{\bibinfo{volume}{76}},
	\bibinfo{pages}{134402} (\bibinfo{year}{2007}).
	
	\bibitem[{\citenamefont{Cheng et~al.}(2018)\citenamefont{Cheng, Shifa, Wang,
			Gao, He, Zhang, Jiang, Liu, and He}}]{Cheng2018}
	\bibinfo{author}{\bibfnamefont{Z.}~\bibnamefont{Cheng}},
	\bibinfo{author}{\bibfnamefont{T.~A.} \bibnamefont{Shifa}},
	\bibinfo{author}{\bibfnamefont{F.}~\bibnamefont{Wang}},
	\bibinfo{author}{\bibfnamefont{Y.}~\bibnamefont{Gao}},
	\bibinfo{author}{\bibfnamefont{P.}~\bibnamefont{He}},
	\bibinfo{author}{\bibfnamefont{K.}~\bibnamefont{Zhang}},
	\bibinfo{author}{\bibfnamefont{C.}~\bibnamefont{Jiang}},
	\bibinfo{author}{\bibfnamefont{Q.}~\bibnamefont{Liu}}, \bibnamefont{and}
	\bibinfo{author}{\bibfnamefont{J.}~\bibnamefont{He}}, \bibinfo{journal}{Adv.
		Mater.} \textbf{\bibinfo{volume}{30}}, \bibinfo{pages}{1707433}
	(\bibinfo{year}{2018}).
	
	\bibitem[{\citenamefont{Zhu et~al.}(2018)\citenamefont{Zhu, Gan, Muhammad,
			Wang, Wu, Liu, Liu, Zhang, He, Jiang et~al.}}]{Zhu2018}
	\bibinfo{author}{\bibfnamefont{W.}~\bibnamefont{Zhu}},
	\bibinfo{author}{\bibfnamefont{W.}~\bibnamefont{Gan}},
	\bibinfo{author}{\bibfnamefont{Z.}~\bibnamefont{Muhammad}},
	\bibinfo{author}{\bibfnamefont{C.}~\bibnamefont{Wang}},
	\bibinfo{author}{\bibfnamefont{C.}~\bibnamefont{Wu}},
	\bibinfo{author}{\bibfnamefont{H.}~\bibnamefont{Liu}},
	\bibinfo{author}{\bibfnamefont{D.}~\bibnamefont{Liu}},
	\bibinfo{author}{\bibfnamefont{K.}~\bibnamefont{Zhang}},
	\bibinfo{author}{\bibfnamefont{Q.}~\bibnamefont{He}},
	\bibinfo{author}{\bibfnamefont{H.}~\bibnamefont{Jiang}},
	\bibnamefont{et~al.}, \bibinfo{journal}{Chem. Commun.}
	\textbf{\bibinfo{volume}{54}}, \bibinfo{pages}{4481} (\bibinfo{year}{2018}).
	
	\bibitem[{\citenamefont{Ouvrard et~al.}(1985)\citenamefont{Ouvrard, Brec, and
			Rouxel}}]{Ouvrard1985}
	\bibinfo{author}{\bibfnamefont{G.}~\bibnamefont{Ouvrard}},
	\bibinfo{author}{\bibfnamefont{R.}~\bibnamefont{Brec}}, \bibnamefont{and}
	\bibinfo{author}{\bibfnamefont{J.}~\bibnamefont{Rouxel}},
	\bibinfo{journal}{Mater. Res. Bulletin} \textbf{\bibinfo{volume}{20}},
	\bibinfo{pages}{1181} (\bibinfo{year}{1985}).
	
	\bibitem[{\citenamefont{Lan\c{c}on et~al.}(2016)\citenamefont{Lan\c{c}on,
			Walker, Ressouche, Ouladdiaf, Rule, McIntyre, Hicks, R\o{}nnow, and
			Wildes}}]{Lancon2016}
	\bibinfo{author}{\bibfnamefont{D.}~\bibnamefont{Lan\c{c}on}},
	\bibinfo{author}{\bibfnamefont{H.~C.} \bibnamefont{Walker}},
	\bibinfo{author}{\bibfnamefont{E.}~\bibnamefont{Ressouche}},
	\bibinfo{author}{\bibfnamefont{B.}~\bibnamefont{Ouladdiaf}},
	\bibinfo{author}{\bibfnamefont{K.~C.} \bibnamefont{Rule}},
	\bibinfo{author}{\bibfnamefont{G.~J.} \bibnamefont{McIntyre}},
	\bibinfo{author}{\bibfnamefont{T.~J.} \bibnamefont{Hicks}},
	\bibinfo{author}{\bibfnamefont{H.~M.} \bibnamefont{R\o{}nnow}},
	\bibnamefont{and} \bibinfo{author}{\bibfnamefont{A.~R.}
		\bibnamefont{Wildes}}, \bibinfo{journal}{Phys. Rev. B}
	\textbf{\bibinfo{volume}{94}}, \bibinfo{pages}{214407}
	(\bibinfo{year}{2016}).
	
	\bibitem[{\citenamefont{Lee et~al.}(2016)\citenamefont{Lee, Lee, Ryoo, Kang,
			Kim, Kim, Park, Park, and Cheong}}]{Lee2016}
	\bibinfo{author}{\bibfnamefont{J.-U.} \bibnamefont{Lee}},
	\bibinfo{author}{\bibfnamefont{S.}~\bibnamefont{Lee}},
	\bibinfo{author}{\bibfnamefont{J.~H.} \bibnamefont{Ryoo}},
	\bibinfo{author}{\bibfnamefont{S.}~\bibnamefont{Kang}},
	\bibinfo{author}{\bibfnamefont{T.~Y.} \bibnamefont{Kim}},
	\bibinfo{author}{\bibfnamefont{P.}~\bibnamefont{Kim}},
	\bibinfo{author}{\bibfnamefont{C.-H.} \bibnamefont{Park}},
	\bibinfo{author}{\bibfnamefont{J.-G.} \bibnamefont{Park}}, \bibnamefont{and}
	\bibinfo{author}{\bibfnamefont{H.}~\bibnamefont{Cheong}},
	\bibinfo{journal}{Nano Lett.} \textbf{\bibinfo{volume}{16}},
	\bibinfo{pages}{7433} (\bibinfo{year}{2016}).
	
	\bibitem[{\citenamefont{Wang et~al.}(2018{\natexlab{b}})\citenamefont{Wang,
			Ying, Zhou, Sun, Wen, Zhou, Li, Zhang, Han, Xiao et~al.}}]{Wang2018}
	\bibinfo{author}{\bibfnamefont{Y.}~\bibnamefont{Wang}},
	\bibinfo{author}{\bibfnamefont{J.}~\bibnamefont{Ying}},
	\bibinfo{author}{\bibfnamefont{Z.}~\bibnamefont{Zhou}},
	\bibinfo{author}{\bibfnamefont{J.}~\bibnamefont{Sun}},
	\bibinfo{author}{\bibfnamefont{T.}~\bibnamefont{Wen}},
	\bibinfo{author}{\bibfnamefont{Y.}~\bibnamefont{Zhou}},
	\bibinfo{author}{\bibfnamefont{N.}~\bibnamefont{Li}},
	\bibinfo{author}{\bibfnamefont{Q.}~\bibnamefont{Zhang}},
	\bibinfo{author}{\bibfnamefont{F.}~\bibnamefont{Han}},
	\bibinfo{author}{\bibfnamefont{Y.}~\bibnamefont{Xiao}}, \bibnamefont{et~al.},
	\bibinfo{journal}{Nature Commun.} \textbf{\bibinfo{volume}{9}},
	\bibinfo{pages}{1914} (\bibinfo{year}{2018}{\natexlab{b}}).
	
	\bibitem[{\citenamefont{Zheng et~al.}(2019)\citenamefont{Zheng, Jiang, Xue,
			Dai, and Feng}}]{Zheng2019}
	\bibinfo{author}{\bibfnamefont{Y.}~\bibnamefont{Zheng}},
	\bibinfo{author}{\bibfnamefont{X.-x.} \bibnamefont{Jiang}},
	\bibinfo{author}{\bibfnamefont{X.-x.} \bibnamefont{Xue}},
	\bibinfo{author}{\bibfnamefont{J.}~\bibnamefont{Dai}}, \bibnamefont{and}
	\bibinfo{author}{\bibfnamefont{Y.}~\bibnamefont{Feng}},
	\bibinfo{journal}{Phys. Rev. B} \textbf{\bibinfo{volume}{100}},
	\bibinfo{pages}{174102} (\bibinfo{year}{2019}).
	
	\bibitem[{\citenamefont{Dovesi et~al.}(2018)\citenamefont{Dovesi, Erba,
			Orlando, Zicovich-Wilson, Civalleri, Maschio, Rérat, Casassa, Baima,
			Salustro et~al.}}]{crystal17}
	\bibinfo{author}{\bibfnamefont{R.}~\bibnamefont{Dovesi}},
	\bibinfo{author}{\bibfnamefont{A.}~\bibnamefont{Erba}},
	\bibinfo{author}{\bibfnamefont{R.}~\bibnamefont{Orlando}},
	\bibinfo{author}{\bibfnamefont{C.~M.} \bibnamefont{Zicovich-Wilson}},
	\bibinfo{author}{\bibfnamefont{B.}~\bibnamefont{Civalleri}},
	\bibinfo{author}{\bibfnamefont{L.}~\bibnamefont{Maschio}},
	\bibinfo{author}{\bibfnamefont{M.}~\bibnamefont{Rérat}},
	\bibinfo{author}{\bibfnamefont{S.}~\bibnamefont{Casassa}},
	\bibinfo{author}{\bibfnamefont{J.}~\bibnamefont{Baima}},
	\bibinfo{author}{\bibfnamefont{S.}~\bibnamefont{Salustro}},
	\bibnamefont{et~al.}, \bibinfo{journal}{WIREs Comput. Mol. Sci.}
	\textbf{\bibinfo{volume}{8}}, \bibinfo{pages}{e1360} (\bibinfo{year}{2018}).
	
	\bibitem[{\citenamefont{Peintinger et~al.}(2013)\citenamefont{Peintinger,
			Oliveira, and Bredow}}]{Peintinger2013}
	\bibinfo{author}{\bibfnamefont{M.~F.} \bibnamefont{Peintinger}},
	\bibinfo{author}{\bibfnamefont{D.~V.} \bibnamefont{Oliveira}},
	\bibnamefont{and} \bibinfo{author}{\bibfnamefont{T.}~\bibnamefont{Bredow}},
	\bibinfo{journal}{J. Comput. Chem.} \textbf{\bibinfo{volume}{34}},
	\bibinfo{pages}{451} (\bibinfo{year}{2013}).
	
	\bibitem[{\citenamefont{Monkhorst and Pack}(1976)}]{Monkhorst1976}
	\bibinfo{author}{\bibfnamefont{H.~J.} \bibnamefont{Monkhorst}}
	\bibnamefont{and} \bibinfo{author}{\bibfnamefont{J.~D.} \bibnamefont{Pack}},
	\bibinfo{journal}{Phys. Rev. B} \textbf{\bibinfo{volume}{13}},
	\bibinfo{pages}{5188} (\bibinfo{year}{1976}).
	
	\bibitem[{\citenamefont{Becke}(1993)}]{B3LYP}
	\bibinfo{author}{\bibfnamefont{A.~D.} \bibnamefont{Becke}},
	\bibinfo{journal}{J. Chem. Phys.} \textbf{\bibinfo{volume}{98}},
	\bibinfo{pages}{5648} (\bibinfo{year}{1993}).
	
	\bibitem[{\citenamefont{Jackson et~al.}(2015)\citenamefont{Jackson, Skelton,
			Hendon, Butler, and Walsh}}]{Jackson2015}
	\bibinfo{author}{\bibfnamefont{A.~J.} \bibnamefont{Jackson}},
	\bibinfo{author}{\bibfnamefont{J.~M.} \bibnamefont{Skelton}},
	\bibinfo{author}{\bibfnamefont{C.~H.} \bibnamefont{Hendon}},
	\bibinfo{author}{\bibfnamefont{K.~T.} \bibnamefont{Butler}},
	\bibnamefont{and} \bibinfo{author}{\bibfnamefont{A.}~\bibnamefont{Walsh}},
	\bibinfo{journal}{J. Chem. Phys.} \textbf{\bibinfo{volume}{143}},
	\bibinfo{pages}{184101} (\bibinfo{year}{2015}).
	
	\bibitem[{\citenamefont{Pascale et~al.}(2004)\citenamefont{Pascale,
			Zicovich-Wilson, L\'{o}pez~Gejo, Civalleri, Orlando, and
			Dovesi}}]{Pascale2004}
	\bibinfo{author}{\bibfnamefont{F.}~\bibnamefont{Pascale}},
	\bibinfo{author}{\bibfnamefont{C.~M.} \bibnamefont{Zicovich-Wilson}},
	\bibinfo{author}{\bibfnamefont{F.}~\bibnamefont{L\'{o}pez~Gejo}},
	\bibinfo{author}{\bibfnamefont{B.}~\bibnamefont{Civalleri}},
	\bibinfo{author}{\bibfnamefont{R.}~\bibnamefont{Orlando}}, \bibnamefont{and}
	\bibinfo{author}{\bibfnamefont{R.}~\bibnamefont{Dovesi}},
	\bibinfo{journal}{J. Comput. Chem.} \textbf{\bibinfo{volume}{25}},
	\bibinfo{pages}{888} (\bibinfo{year}{2004}).
	
	\bibitem[{\citenamefont{Bernasconi et~al.}(1988)\citenamefont{Bernasconi,
			Marra, Benedek, Miglio, Jouanne, Julien, Scagliotti, and
			Balkanski}}]{Bernasconi1988}
	\bibinfo{author}{\bibfnamefont{M.}~\bibnamefont{Bernasconi}},
	\bibinfo{author}{\bibfnamefont{G.~L.} \bibnamefont{Marra}},
	\bibinfo{author}{\bibfnamefont{G.}~\bibnamefont{Benedek}},
	\bibinfo{author}{\bibfnamefont{L.}~\bibnamefont{Miglio}},
	\bibinfo{author}{\bibfnamefont{M.}~\bibnamefont{Jouanne}},
	\bibinfo{author}{\bibfnamefont{C.}~\bibnamefont{Julien}},
	\bibinfo{author}{\bibfnamefont{M.}~\bibnamefont{Scagliotti}},
	\bibnamefont{and}
	\bibinfo{author}{\bibfnamefont{M.}~\bibnamefont{Balkanski}},
	\bibinfo{journal}{Phys. Rev. B} \textbf{\bibinfo{volume}{38}},
	\bibinfo{pages}{12089} (\bibinfo{year}{1988}).
	
	\bibitem[{\citenamefont{Wang et~al.}(2016)\citenamefont{Wang, Du, Liu, Hu,
			Zhang, Zhang, Owen, Lu, Gan, Sengupta et~al.}}]{Wang2016}
	\bibinfo{author}{\bibfnamefont{X.}~\bibnamefont{Wang}},
	\bibinfo{author}{\bibfnamefont{K.}~\bibnamefont{Du}},
	\bibinfo{author}{\bibfnamefont{Y.~Y.~F.} \bibnamefont{Liu}},
	\bibinfo{author}{\bibfnamefont{P.}~\bibnamefont{Hu}},
	\bibinfo{author}{\bibfnamefont{J.}~\bibnamefont{Zhang}},
	\bibinfo{author}{\bibfnamefont{Q.}~\bibnamefont{Zhang}},
	\bibinfo{author}{\bibfnamefont{M.~H.~S.} \bibnamefont{Owen}},
	\bibinfo{author}{\bibfnamefont{X.}~\bibnamefont{Lu}},
	\bibinfo{author}{\bibfnamefont{C.~K.} \bibnamefont{Gan}},
	\bibinfo{author}{\bibfnamefont{P.}~\bibnamefont{Sengupta}},
	\bibnamefont{et~al.}, \bibinfo{journal}{2D Materials}
	\textbf{\bibinfo{volume}{3}}, \bibinfo{pages}{031009} (\bibinfo{year}{2016}).
	
	\bibitem[{\citenamefont{Momma and Izumi}(2011)}]{VESTA}
	\bibinfo{author}{\bibfnamefont{K.}~\bibnamefont{Momma}} \bibnamefont{and}
	\bibinfo{author}{\bibfnamefont{F.}~\bibnamefont{Izumi}}, \bibinfo{journal}{J.
		Appl. Crystallogr.} \textbf{\bibinfo{volume}{44}}, \bibinfo{pages}{1272}
	(\bibinfo{year}{2011}).
	
\end{thebibliography}
\end{document}